# Strongly Enhanced Current-Carrying Performance in MgB$_2$ Tape Conductors by Novel C$_{60}$ Doping

Xianping Zhang, Yanwei Ma[*], Zhaoshun Gao, Dongliang Wang, Lei Wang

Key Laboratory of Applied Superconductivity, Institute of Electrical Engineering, Chinese Academy of Sciences, P.O. Box 2703, Beijing 100080, China

Wei Liu, Chunru Wang

Institute of Chemistry, Chinese Academy of Sciences, Beijing 100080, China

**Abstract:**

MgB$_2$ is a promising superconductor for important large-scale applications for both high field magnets and cryocooler-cooled magnet operated at temperatures around 20 K. In this work, by utilizing C$_{60}$ as a viable alternative dopant, we demonstrate a simple and industrially scaleable rout that yields a 10~15-fold improvement in the in-high-field current densities of MgB$_2$ tape conductors. For example, a $J_c$ value higher than $4\times10^4$ A/cm$^2$ (4.2 K, 10 T), which exceeds that for NbTi superconductor, can be realized on the C$_{60}$ doped MgB$_2$ tapes. It is worth noting that this value is even higher than that fabricated using strict high energy ball milling technique under Ar atmosphere. At 20 K, $H_{irr}$ was ~10 T for C$_{60}$ doped MgB$_2$ tapes. A large amount of nanometer-sized precipitates and grain boundaries were found in MgB$_2$ matrix. The special physical and chemical characteristic of C$_{60}$, in addition to its C containing intrinsic essence, is a key point in enhancing the superconducting performance of MgB$_2$ tapes.

---

[*] Author to whom correspondence should be addressed, E-mail: ywma@mail.iee.ac.cn

The MgB$_2$ superconductor has aroused great research interest in the area of fundamental and applied research [1-4] because of its rather high transition temperature ($T_c$), two-gap superconductivity, lacking weak links, and low material cost. It is a promising candidate for applications such as magnetic resonance imaging, high-field magnets, transmission cables, and current leads. Actually, some engineering program have been carried out based on MgB$_2$ superconductor.[5] However, critical current density ($J_c$) of MgB$_2$ under magnetic fields is still lower than that for practical uses. Various synthesis techniques have been developed to improve the $J_c$-B properties of MgB$_2$.[6-9] But for the fabrication of long length MgB$_2$ conductors required by large-scale superconducting magnet applications, element/compound doping process is the most commonly used procedure. Among the large number of dopants reported, C, SiC, hydrocarbons, etc., are the most favorable additives, being effective for improvement of the upper critical field ($H_{c2}$) and $J_c$ in magnetic fields.[10-13] When MgB$_2$ were doped with these materials, C substitutes for B and introduces electron scattering and impurity scattering which reduce the mean free path and the coherence length ξ, increasing $H_{c2}$.[14] It has been reported that $H^{//}_{c2}$(4.2 K) in C doped thin films have reached 51 T.[15] However, the doping effect to MgB$_2$ superconductor has been restrained by conglomeration of the dopants or/and by poor reactivity between MgB$_2$ and dopants. For instance, the $H_{c2}$ and $J_c$ in nano-C doped samples can be obviously improved only when the samples was sintered at temperature up to 900 °C, while SiC doping usually results in the agglomerations of impurity phases. Hydrocarbons are investigated hotly these days, but they are not suitable to fabricate long length MgB$_2$ superconductor because of the gas generated.[2] Therefore, it is necessary to investigate new dopant which is suitable for the fabrication of high performance long length MgB$_2$ conductor.

Here we report a novel MgB$_2$ dopant, C$_{60}$, which can ensure the homogeneous dispersion of the impurities and efficient substitution of B by C. Fullerene C$_{60}$ is a remarkable class of molecules in which large numbers of carbon molecules are locked together into a roughly spherical shape.[16, 17] Each C$_{60}$ molecule consists of 60 carbon atoms, arranged as 12 pentagons and 20 hexagons. The C$_{60}$ molecule is very small (~0.71 nm), and $C_{60}$ molecules condense to form a solid of weakly bound molecules. In some case the $C_{60}$ molecule geometrical configuration can be broken and solo C atoms will be produced. The small molecule of $C_{60}$ gives a highly uniform mixture, while the high reactivity of freshly C substitutes B at low temperature. Therefore, $C_{60}$ is an ideal dopant for $MgB_2$ composites to improve its $J_c$-$B$ performance.

In this work, it is demonstrated that $C_{60}$ doping resulted in a significant enhancement of the $J_c$-$B$ performance over the entire applied high-field range. From the analysis of $MgB_2$ lattice parameter, the C substitution is actually happened in $C_{60}$ doped $MgB_2$ tapes. Compared to nano-C doped samples, the sintering temperature is much lower to obtain the same C substitution level. The $J_c$ value is much higher in $C_{60}$ doped tapes than that of nano-C doped $MgB_2$ samples with similar irreversibility field ($H_{irr}$), showing different intrinsic property in these two type samples.

$C_{60}$-doped $MgB_2$ tapes were fabricated by the commonly reported powder-in-tube (PIT) method. Commercial powders of magnesium (-325 mesh, 99.8%), boron (amorphous, 2-5 $\mu$m, 99.99%) and $C_{60}$ (99%) were weighed out according to the nominal atomic ratio of $(Mg_{1.05}B_2)_{1-x}(C)_x$ with x = 0, 0.05, 0.08, 0.10, 0.15, respectively. When they were mixed in air for an hour, the powders were put into pure iron tubes with an outer diameter of 8 mm, and an inner diameter of 5 mm. After packing, the tubes were rotary-swaged and drawn to wires of 1.75 mm in diameter. The wires were subsequently rolled into flat tapes. It should be noted that all the tape processing procedures were carried out in the air condition. The final tapes were cut and raped in Ta foil, then sintered at 700-900°C for 1 h in flowing high purity Ar, followed by a furnace cooling to room temperature.

The phase and crystal structure of all the samples were investigated using powder x-ray diffraction (XRD) with Cu $K\alpha$ radiation. The lattice parameters were obtained from the analysis of the diffraction data using the X`pert program. The resistance measurements were carried out on Physical Properties Measurement System (PPMS) up to 9 T using small pieces of $MgB_2$ core after peeling off the Fe sheath or on a 20 T magnet using Fe sheathed $MgB_2$ tapes.[22] The transport $I_c$ at 4.2 K was measured by the four-point-probe resistive method with a criterion of 1 $\mu$V cm$^{-1}$ at the High Field

Laboratory For Superconducting materials (HFLSM) at Sendai in Japan. The grain morphology and microstructure were examined by transmission electron microscopy (TEM) equipped with an x-ray energy dispersion spectrum (EDS).

Figure 1 shows XRD patterns of $MgB_2$ samples sintered at 800 °C with different $C_{60}$ doping level. It can be seen that all of the samples exhibit a well-developed $MgB_2$ phase. The diffraction peaks MgO are relatively high in our samples, making it not easy to accurately calculate the lattice parameters of $MgB_2$. On the other hand, the (110) and (100) peaks of doped samples were shifted obviously to higher angles compared to the undoped ones, while the position of (002) peak shows no obvious shift. This means $c$-axis lattice parameters did not vary significantly within the set, but $a$-axis lattice parameters decreased when $C_{60}$ were added. The decrease of the $a$-axis is an indication of the carbon substitution for boron,[18] which is further proved by the depression of the critical transition temperature ($T_c$) of the $C_{60}$ doped samples. The C substitution to B site has a great impact on the carrier density and impurity scattering. It is expected that carbon, which has one more electron than boron, will donate electrons to the σ band. Also, an increase of impurity scattering within the π band and the modification of band structure can be achieved by carbon substitution. At the same time, x-ray lines are broadened in doped samples, indicating a degraded crystallinity. This may be caused by small crystallite or domain size and lattice defects.[19] It is well known that various types of lattice distortion, intragranular precipitates and small crystal sizes, which increase the intraband scattering in both σ and π bands and shorten the electron mean free path and coherence length, usually result in an improvements on $H_{c2}$ and $H_{irr}$.[9] Therefore, high $H_{c2}$ and $H_{irr}$ are expected in $C_{60}$ doped samples.

Table I provides an overview of the properties of $MgB_2$ samples doped with $C_{60}$. Clearly, $C_{60}$ additions increased residual resistivity $\rho(40K)$, and decreased residual resistivity ratio RRR=$\rho(300K)/\rho(40K)$. The data for undoped sample display the usual properties of pristine $MgB_2$ with a low ratio of the normal sate and room-temperature resistivity. For the $C_{60}$ doped samples, there is a remarkable increase of the $\rho(40K)$. The residual resistivity increase may be partly explained by the nonsuperconducting and generally insulating second phases present in the doped samples.[20] Another reason is likely due to an enhancement in intraband scattering,[5] induced by C substitution, which causes a reduction in the electron mean free path.

The residual resistivity ratio, $\rho(300\ \text{K})/\rho(40\ \text{K})$ for the $C_{60}$ doped samples showed a drastic decrease compared to undoped samples. It decreases from 2.15 in pure samples to 1.54 in the 8% $C_{60}$ doped samples and to 1.31 in the 10% $C_{60}$ doped ones. This is typical for $MgB_2$ samples with a high concentration of defects or impurities. Enhanced scattering at the grain boundaries contributes to the decrease of the RRR values. The active area fraction ($A_F$) is calculated using a modified version of Rowell's formalism, $A_F = \Delta\rho_{ideal}/[\rho(300\ \text{K})-\rho(40\ \text{K})]$, where $\Delta\rho_{ideal} = 7.3\ \mu\Omega$ cm.[21] As listed in Table I, $A_F$ decreased from 0.23 for the undoped samples to 0.08 for 10% $C_{60}$ doped. The impurity phases existed at the grain boundary, which decrease the superconducting volume fraction and current flow channel, may be accounted for this.[20]

Figure 2 shows the temperature dependence of $H_{irr}$ and $H_{c2}$ for the pure $MgB_2$ tapes and $C_{60}$ doped samples, where the $H_{c2}$ and $H_{irr}$ obtained from the 90% and 10% values of the normal-state resistance $\rho(40K)$ measured using PPMS with voltage and current taps sticking on $MgB_2$ core. Clearly, as a result of $C_{60}$ doping, the $H_{c2}$ ($H_{irr}$)-$T$ curve became steeper, indicating an improved $H_{c2}$ and $H_{irr}$.[22] For instance, at 20 K, $H_{irr}$ was nearly 10 T for $C_{60}$ doped $MgB_2$ tapes sintered at 800 °C. This is in accordance with XRD analysis that C has been substituted into $MgB_2$ lattice. C in $MgB_2$ lattice can act as a point defect, enhancing scattering primarily within the $\pi$ band, and leading to an improvement of $H_{c2}$ at low doping levels.[23] The inset of figure 2 exhibits the temperature dependence of $H_{irr}$ obtained from the 10% values of the normal-state resistance $\rho(40K)$ measured on a 20 T magnet, with voltage and current taps directly soldered on Fe sheath of the $MgB_2$/Fe tapes. Although the $H_{irr}$ value is a little lower than that measured by PPMS due to different sample preparation for measurements,[10] it shows a similar improvement by $C_{60}$ doping. At 27.5 K, there is a cross for $H_{irr}$-temperature lines of $C_{60}$ doped samples and nano-C doped samples,[24] indicating a higher $H_{irr}$ value at lower temperature in $C_{60}$ doped samples.

Transport measurements at 4.2 K in fields up to 14 T were conducted for $C_{60}$ doped and undoped samples sintered at 700 and 800 °C. As can be seen from figure 3, very high $J_c$ values were obtained even sintered at low temperature. For examples, a $J_c$ value as high as $3.6\times10^4$ A/cm$^2$ (4.2 K, 10 T) was reached in the 10% $C_{60}$ doped $MgB_2$ tapes sintered at 700 °C. This value not only exceeds that obtained in nano-C doped samples sintered at 900 °C,[24] but also shows much higher than nano-SiC doped samples.[11] In particular, at 11 T (4.2 K), a very high $J_c$ value, which is about $2.8\times10^4$ A/cm$^2$, was obtained for the 10% $C_{60}$ doped samples sintered at 800 °C. Referring to the best data of Herrmann[25] and Yamada et al.,[26] whose C or SiC doped tapes were made by using strict high energy ball milling technique under Ar atmosphere, this very high $J_c$ is the highest of reported values for $MgB_2$ tapes fabricated by common PIT method so far. Moreover, the $I_c$ was too high to be measured below 11 T for samples heated at 800°C because the quenching behavior occurred during the $I_c$ measurements. However, the extrapolated $J_c$ line to low magnetic fields reached a value higher than $4\times10^4$ A/cm$^2$ at 4.2 K, 10 T. This is a landmark value which is higher than that for NbTi superconductor at the same circumstance,[27] with an outstanding $J_c$-$B$ property in high fields, as shown in figure 3. The $C_{60}$ doping in $MgB_2$ opens up an effective and easily controlled method to improve Jc. It should be noted that this technique is very suitable for the industrial scale fabrication of $MgB_2$ tapes and wires because the $C_{60}$-doped wire samples are prepared in the air condition.

Considering the extremely high $J_c$ value even at low sintering temperature, and the relatively low carbon substitution level, we think the large amount of grain boundaries and nanometer-sized precipitates play an important role, as will be discussed below. On the other hand, with increasing the sintering temperature, the $J_c$ values were further improved. The improvement of the $H_{irr}$ may be the main reason. At higher sintering temperature, more C and/or nanometer-sized precipitates are incorporated into $MgB_2$ matrix, which in turn bring more lattice distortion, acting as effective pinning centers to improve the pinning ability, quite similar to the data of Yeoh et al.[12] Unfortunately, it was not possible to measure the transport $J_c$ of the 10% $C_{60}$ doped samples sintered at 900 °C due to the serious slope of the I-V line, which is caused by the high resistivity of reaction layer between Fe and $MgB_2$.

In order to understand the mechanisms for the enhancement of $J_c$ at higher fields in $C_{60}$ doped samples, a TEM study was performed. Figure 4 shows the typical low magnification and high-resolution TEM micrographs for 10% $C_{60}$ doped samples sintered at 700 °C (Fig. 4 (a) and (b)) and 800 °C (Fig. 4 (c) and (d)). The selected area diffraction (SAD) patterns, as shown in the corner of the Fig. 4 (a) and (c), consist of well defined ring patterns, meaning a very fine grain size.[4] TEM images showed that $MgB_2$ grains in some areas is hard to distinguish as the grains are so well consolidated, and there are a lager number of nanometer-sized precipitates (<10 nm) scattered throughout the $MgB_2$ grains, as pointed out by arrows in figure 4. These defects can cause strong scattering and enhance both $J_c$ and $H_{c2}$. EDS analysis on a large portion of the $MgB_2$ grain revealed that Mg, B, O, C coexist in these areas. Based on the EDS results and the shrinkage of lattice parameter $a$ in $C_{60}$ doped samples, we suggests that C were actually substituted into $MgB_2$ lattice. The high density of grain boundaries induced by small $MgB_2$ grain size and nanometer-sized precipitates scattered in $MgB_2$ grains can served as strong pinning centers. Accordingly, the high $J_c$-$B$ property of $C_{60}$ doped samples may be attributed to these large amounts of defects and grain boundaries.

As discussed in previous literatures,[3,10,24] for C and SiC doped samples, the high $J_c$ value in magnetic fields are attributed to the smaller grain size, strong vortex pinning ability, and high $H_{c2}$ induced by C substitution. Even though this is nearly the same phenomenon in $C_{60}$ doped samples, the $J_c$-$B$ properties of $C_{60}$ doped $MgB_2$ is much higher than that in C and SiC doped samples made by the same fabrication process. As discussed by Yeoh,[12] the major improvement of $H_{c2}$ and $J_c$ for nano-C and SiC doped $MgB_2$ has a different origin. But in $C_{60}$ doped samples, it seems that grain boundary defects and carbon substitution are coexisting and playing an important factor in the enhancement of $H_{c2}$ and $J_c$. The special characteristic of $C_{60}$ is thought to be accounted for this, as there is no other phase in staring $C_{60}$ powder (seen from the XRD pattern in Fig. 5). The structure of $C_{60}$ fullerene has two different types of bonds one at the junction of two hexagons or 6-6 junctions and the other at the junction of pentagon and hexagon or 5-6 junctions. With this geometrical configuration, the $C_{60}$ molecule is very small, and they can act as favorable pining centers in $MgB_2$ grains. At the same time, they can serve as nucleation sites for grain formation and obtain very fine $MgB_2$ grains. Moreover, $C_{60}$ will become superconductors ($T_c$: 20-40 K) if certain alkali atoms (for example, K or Rb) are added to solid $C_{60}$.[28] It is still not clear if this character play some role in the improvement of the $J_c$-$B$ properties of $MgB_2$ tapes.

As mentioned earlier, the sizes of Mg and B particles used are very large, and there are many MgO formed in our samples. These two factors strongly affect the final $J_c$-$B$ property of $MgB_2$ tapes.[13, 29] But there is a solution to solve these problems, for instance, better $J_c$-$B$ properties have been achieved in undoped and C doped $MgB_2$ tapes using high energy ball milling technique under Ar atmosphere.[25] If this fabrication technique is applied to $C_{60}$ doped samples, a much higher $J_c$ value is expected. As we know, $C_{60}$ has been commercialized at present, so it is a very promising $MgB_2$ dopant for practical uses. New phenomenon may also be found based on the unusual electrical, mechanical, and thermal properties of $C_{60}$ and two-gap structure of $MgB_2$.

In summary, a more favorable $MgB_2$ dopant, $C_{60}$, was developed. The $J_c$-$B$ property enhancement of $C_{60}$ doped $MgB_2$ tapes is much higher than that by the commonly used C and SiC doping. Besides its C containing intrinsic property, the very small molecule and special chemical characteristic of $C_{60}$ are believed to be the main reasons. With these characteristics, $C_{60}$ enables homogeneous distribution of defects in $MgB_2$ crystal lattice and effective substitution of boron atoms by carbon during the solid-state reaction process. This work open up a new direction for the research of $MgB_2$ dopant, which have been stagnated for several years. More importantly, with the great improvement in the $J_c$-$B$ properties by $C_{60}$ doping using a relatively sample technique, the practical large scale uses of $MgB_2$ superconductor is going to become true.

*Acknowledgment*

We thank Prof. Kazuo Watanabe, Haihu Wen, Xuedong Bai, Liye Xiao and Liangzhen Lin for their help and useful discussions. This work is partially supported by the Beijing municipal science and technology commission (grant no. Z07000300703), National '973' Program (grant no. 2006CB601004) and National '863' Project (grant no. 2006AA03Z203).

TABLE I. Sample data for $MgB_2$ tapes made by different sintering temperatures and $C_{60}$ doping levels.

| Doping level (at.%) | Sintering temperature (°C) | Lattice $a$ (Å) | FWHM (101) (°) | $\rho_{40K}$ (μΩ-cm) | RRR | $A_F$ | $H_{irr}$ (T) (20 K) | $J_c$ (A/cm$^2$) (4.2K,10T) |
|---|---|---|---|---|---|---|---|---|
| 0 | 800 | 3.0829 | 0.478 | 30.9 | 1.95 | 0.21 | 7.3 | 3.5×10$^3$ |
| 5 | 800 | - | 0.771 | 173 | 1.37 | 0.11 | - | 2.1×10$^4$ |
| 8 | 800 | - | 0.669 | 210 | 1.54 | 0.06 | 8.9 | 2.4×10$^4$ |
| 10 | 800 | 3.0760 | 0.673 | 307 | 1.31 | 0.08 | > 9 | >4.0×10$^4$ |
| 10 | 700 | 3.0776 | 0.717 | 406 | 1.39 | 0.05 | 8.3 | 3.6×10$^4$ |
| 10 | 900 | 3.0679 | 0.628 | 296 | 1.33 | 0.07 | - | - |

**Captions**

Figure 1. XRD patterns of MgB$_2$ samples doped with different C$_{60}$ ratio. The samples were sintered at 800 °C for 1 h. The peaks of MgB$_2$ indexed, while the peaks of MgO are marked by asterisks.

Figure 2. The temperature dependence of $H_{irr}$ and $H_{c2}$ for undoped and 10% C$_{60}$ doped samples sintered at 800 °C. The temperature dependence of $H_{irr}$ for 10% C$_{60}$ doped and undoped samples measured with Fe sheath were plotted as inset.

Figure 3. $J_c$-$B$ properties of undoped and C doped tapes heated at 700 and 800 °C.

Figure 4. TEM micrographs showing the nanoparticle inclusions and dislocations of the C$_{60}$ doped samples. The inset displays the selected-area electron diffraction pattern taken from the circular region of 200 nm in diameter

Figure 5. TEM image (a) and XRD pattern (b) of the C$_{60}$ starting powder, with a high-resolution TEM micrograph was put at the corner of TEM image.

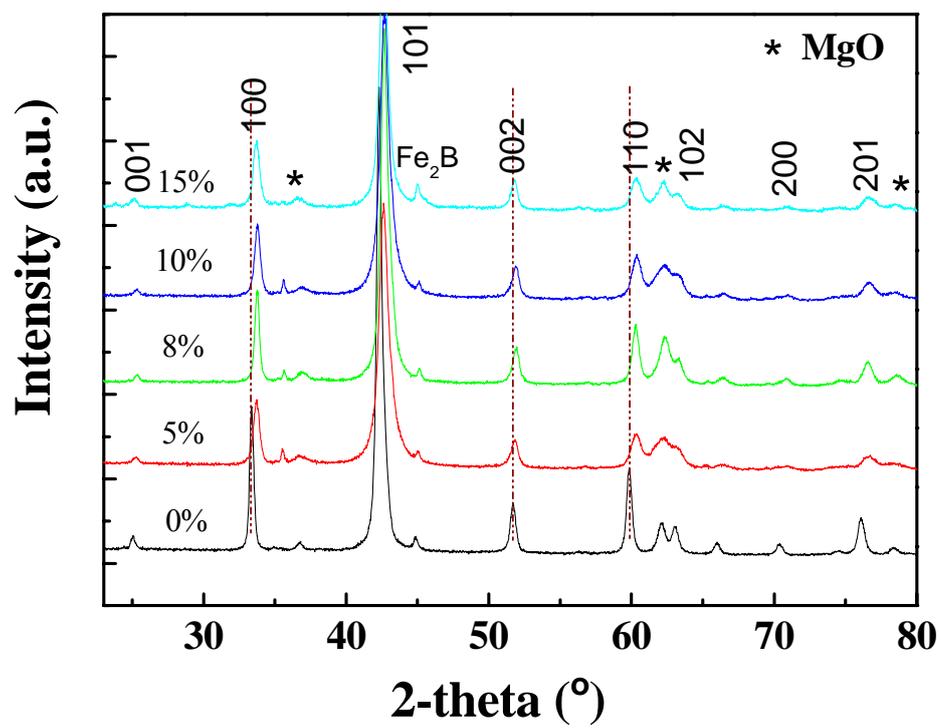

Fig.1 Zhang et al.

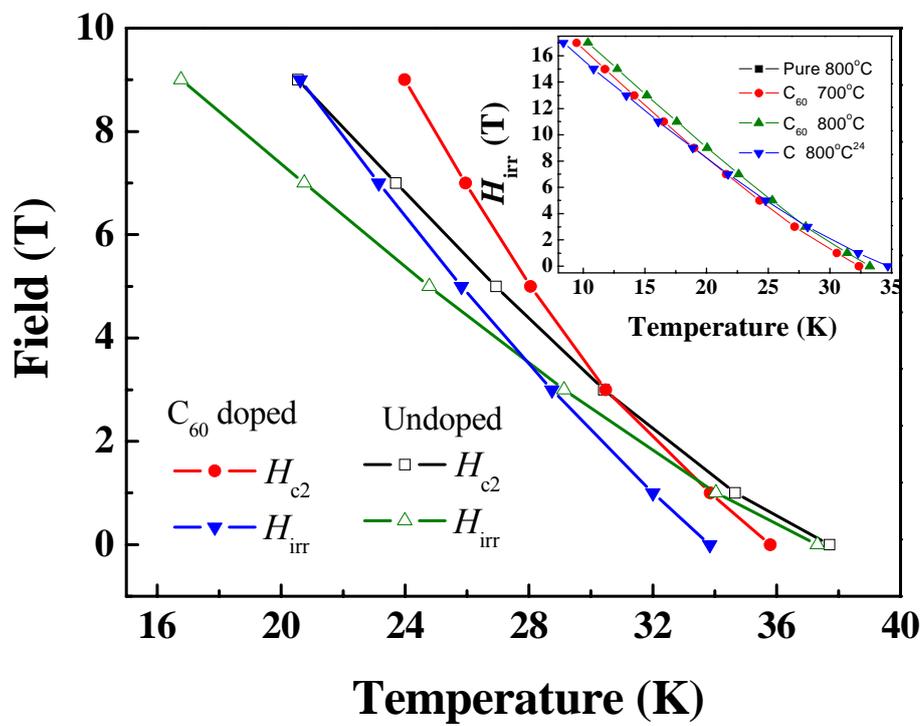

Fig.2 Zhang et al.

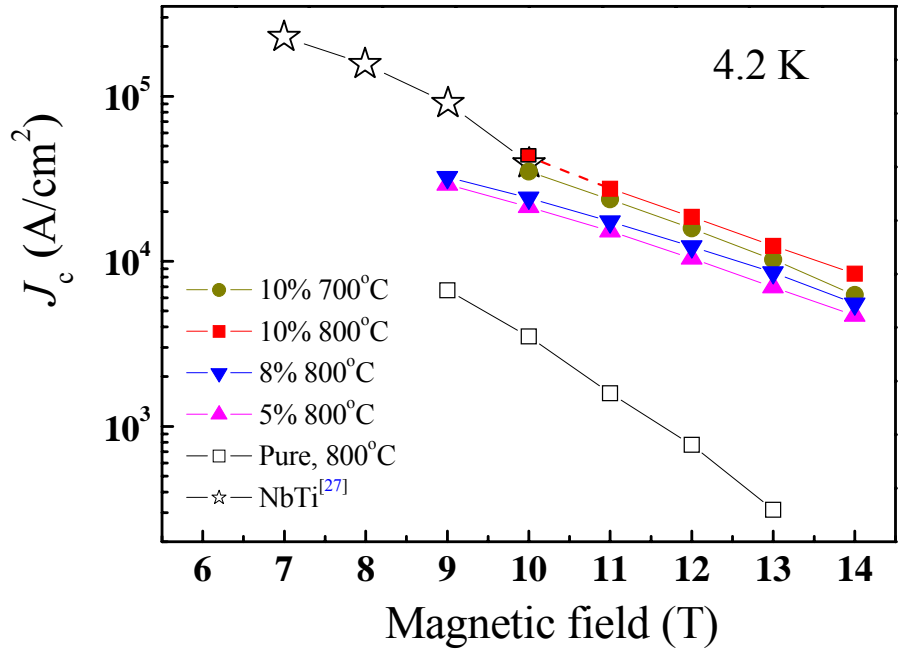

Fig.3 Zhang et al.

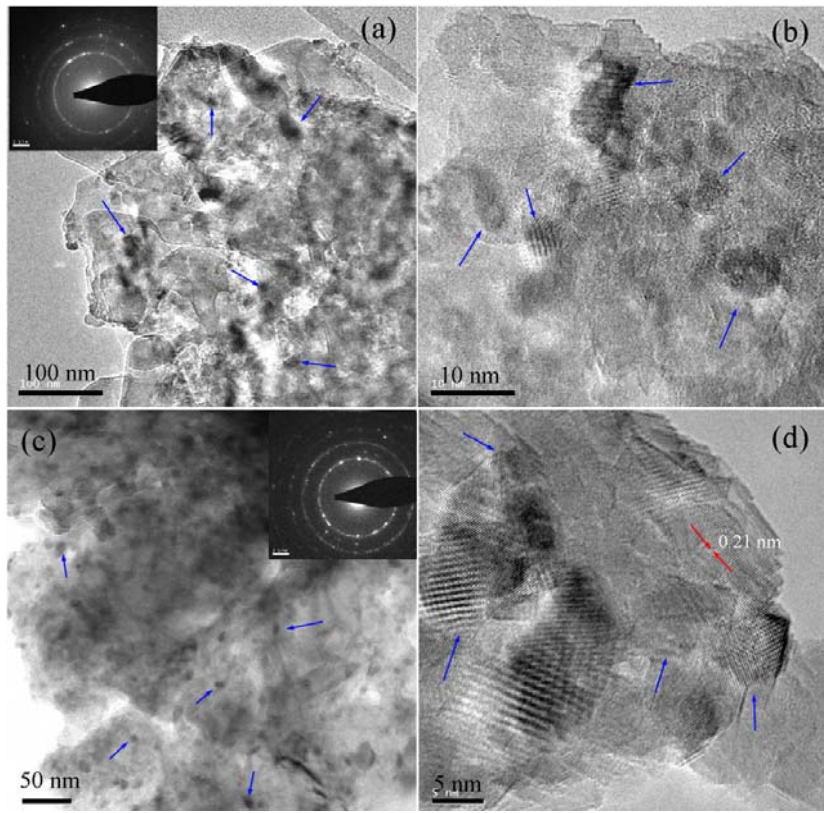

Fig.4 Zhang et al.

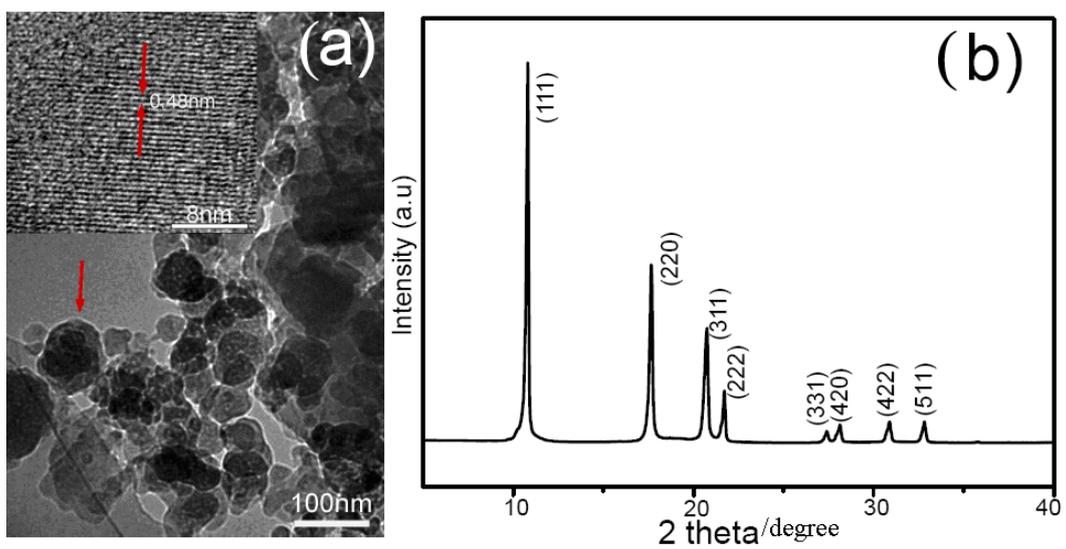

Fig.5 Zhang et al.